\documentclass[12pt]{article}
\usepackage{amsfonts}

\usepackage{amsmath}
\usepackage{cite}
\usepackage{graphicx}

\csname @addtoreset\endcsname{equation}{section}
\newcommand{\eq}{\begin{equation}}
\newcommand{\feq}{\end{equation}}
\newcommand{\eqn}{\begin{eqnarray}}
\newcommand{\feqn}{\end{eqnarray}}
\newcommand{\arr}{\begin{eqnarray*}}
\newcommand{\farr}{\end{eqnarray*}}

\input{tcilatex}

\begin{document}

\begin{titlepage}
\begin{flushright}
CAMS/02-05\\
\end{flushright}
\vspace{.3cm}
\begin{center}
\renewcommand{\thefootnote}{\fnsymbol{footnote}}
{\Large \bf Curved Branes in AdS Einstein-Maxwell Gravity and Killing Spinors}
\vfill
{\large \bf { W.~A.~Sabra\footnote{email: ws00@aub.edu.lb}}}\\
\renewcommand{\thefootnote}{\arabic{footnote}}
\setcounter{footnote}{0}
\vfill
{Center for Advanced Mathematical Sciences (CAMS)\\
and\\
Physics Department, American University of Beirut, Lebanon.}
\end{center}
\vfill
\begin{center}
{\bf Abstract}
\end{center}

We determine the Killing spinors for a class of magnetic brane solutions with Minkowski
worldvolume of the theory of AdS Einstein Maxwell theories in $d$ dimensions.
We also obtain curved magnetic brane solutions with Ricci-flat worldvolumes.
If we demand that the curved brane solution admits Killing spinors, then
its worldvolume must admit parallel spinors.
Classes of Ricci-flat worldvolumes admitting parallel spinors are discussed.
\end{titlepage}

\section{Introduction}

The conjectured equivalence between string theory on anti-de~Sitter (AdS)
space and certain superconformal gauge theories living on the boundary of
this space \cite{ads} has led to a lot of interest in the study of AdS
gravitational configurations. In part, this interest has also been motivated
by the suggestion that four dimensional Einstein gravity can be recovered on
a domain wall embedded in AdS space \cite{RS}. Classical AdS solutions may
provide the possibility to study the nonperturbative structure of the field
theories living on the boundary.

Black hole solutions to Einstein's equations with a negative cosmological
constant were constructed many years ago in \cite{carter}. Some activity has
been recently devoted to the study of solutions with various horizon
topologies given by planar, toroidal or arbitrary genus Riemann surface \cite
{Lemos, Hu, Ca, Be, Mann, Vanzo, Brill, rot, Birm}. In the context of gauged
supergravity models, black hole and brane solutions were discussed in many
places (see for example \cite{romans, ck, cs,london, perr, klemm, bcs2, csb,
kkk, chamsabra, ks, duff, dufftasi, cveduff, gaun, liu2}).

In recent years certain supersymmetric curved $p$-brane and domain wall
solutions were constructed \cite{perry, janssen, aliwaf}. A typical $p$%
-brane solution is a warped product of two spacetimes, one describing a $%
(p+1)$-dimensional Minkowski worldvolume and the other being the transverse
space to the brane. For the solutions discussed in \cite{perry, janssen,
aliwaf}, the Einstein's equations of motion were shown to be satisfied also
when the worldvolume metric is Ricci-flat. The curved solution obtained
admits Killing spinors provided that its Ricci-flat worldvolume admits
parallel spinors. The classification of possible static brane worldvolumes
is related to the known classification of Riemannian holonomy groups \cite
{perry}. Static metrics admitting parallel spinors are automatically
Ricci-flat. However, if one allows for indecomposable Ricci-flat Lorentzian
worldvolumes with parallel spinors, more curved solutions with Killing
spinors can be obtained \cite{jose}. Lorentzian manifolds admitting parallel
spinors need not be Ricci-flat. This means that restricting the holonomy of
a given manifold (for it to have parallel spinors) is not enough for its
metric to satisfy Einstein's equations of motion; more restrictions must be
imposed.

In this work we are mainly interested in the study of a class of magnetic
brane solutions with various horizon topologies in $d$-dimensional gravity
with a negative cosmological constant coupled to an abelian gauge field. In 
\cite{waf} $d$-dimensional magnetic brane solutions with flat, hyperbolic
and spherical transverse space and with Minkowski worldvolume were derived.
Motivated by the results of \cite{chamsabra, ks}, we show that these
solutions can be obtained by searching for magnetically charged
configurations with Killing spinors. We determine the projection conditions
on the spinors and their explicit forms. In the context of gauged
supergravity theories, solutions with Killing spinors are supersymmetric. We
also generalize the solutions of \cite{waf} and discover new $d$-dimensional
magnetic brane solutions with curved Ricci-flat worldvolumes.

\section{Magnetic Branes with Killing Spinors}

In this section we obtain magnetic brane solutions of the theory of
Einstein-Maxwell theory with a negative cosmological constant in dimensions $%
d\geq 5$. This is done by searching for magnetically charged brane
configurations admitting Killing spinors. The Lagrangian of our theory can
be given by 
\begin{equation}
e^{-1}\mathcal{L}=R-\Lambda -\frac{(d-2)}{2(d-3)}F_{\mu \nu }F^{\mu \nu }
\label{action}
\end{equation}
where the cosmological constant is $\Lambda =-(d-1)(d-2)l^{2},$ $R$ is the
scalar curvature and $F_{\mu \nu }$ is the field strength of an abelian
gauge field $A_{\mu }.$ Motivated by the equations of gauged supergravity,
we write as a Killing spinor equation in $d$ dimensions 
\begin{equation}
\left[ \mathcal{D}_{\mu }+{\frac{i}{4(d-3)}}\left( \gamma _{\mu }{}^{\nu
\rho }-2(d-3)\delta _{\mu }{}^{\nu }\gamma ^{\rho }\right) F_{\nu \rho }{}-%
\frac{i\left( d-2\right) l}{2}A_{\mu }+\frac{l}{2}\gamma _{\mu }\right]
\epsilon =0.  \label{st}
\end{equation}
Here $\mathcal{{D}_{\mu }}$ is the covariant derivative given by $\mathcal{D}%
_{\mu }=\partial _{\mu }+{\ \frac{1}{4}}\omega _{\mu ab}\gamma ^{ab},$ where 
$\omega _{\mu ab}$ is the spin connection and $\gamma _{a}$ are Dirac
matrices. In this work, we use the metric $\eta ^{ab}=(-,+,+,+,+,....)$ and $%
\{{\gamma ^{a},\gamma ^{b}}\}=2\eta ^{ab}$. In supergravity theories, the
strategy for finding supersymmetric bosonic solutions is as follows. The
fermi fields are set to zero leading automatically to the vanishing of the
supersymmetry variations of the bosonic fields of the theory. The conditions
for the existence of solutions preserving some supersymmetry are then
obtained from the vanishing of the supersymmetry variation of the fermi
fields. AdS Einstein-Maxwell theories with spacetime dimensions $d<6$ can be
obtained as consistent truncations of gauged supergravity models and (\ref
{st}) comes from the vanishing of the gravitini supersymmetry transformation
in a bosonic background. As in \cite{waf}, we consider a general $(d-4)$%
-magnetic brane solution with metric given in the form 
\begin{equation}
ds^{2}=e^{2V(r)}\left( -dt^{2}+dz^{\alpha _{i}}dz^{\alpha _{i}}\right)
+e^{2U(r)}dr^{2}+N^{2}d\Omega ^{2}.
\end{equation}
In what follows we will only consider either $N(r)=r$ or $N(r)=\mathcal{N}=$
constant. Moreover, $d\Omega ^{2}$ denotes the metric of a two-manifold $%
\mathcal{S}$ of constant Gaussian curvature $k$ and we choose 
\begin{equation}
d\Omega ^{2}=d\theta ^{2}+f^{2}d\phi ^{2},
\end{equation}
where 
\begin{equation}
f=\left\{ 
\begin{array}{l@{\quad,\quad}l}
\sin \theta  & k=1, \\ 
1 & k=0, \\ 
\sinh \theta  & k=-1.
\end{array}
\right. 
\end{equation}
$\mathcal{S}$ is a quotient space of the universal coverings $S^{2}$ ($k=1$%
), $H^{2}$ ($k=-1$) or $E^{2}$ ($k=0$). Our branes carry magnetic charge and
therefore our gauge field is given by 
\begin{equation}
F_{\theta \phi }=\pm kqf,\qquad A_{\phi }=\pm kq\int fd\theta .
\end{equation}
Returning to our metric, the connection forms are given by

\begin{equation}
w_{\text{ }1}^{0}=V^{\prime }e^{V-U}\text{ }dt,\text{ \ }w_{\text{ }%
1}^{\alpha _{i}}=V^{\prime }e^{V-U}\text{ }dz^{\alpha _{i}},\text{ \ }w_{%
\text{ }1}^{2}=N^{\prime }e^{-U}d\theta ,\text{ \ }w_{\text{ }%
1}^{3}=fN^{\prime }e^{-U}d\phi ,\text{ \ }w_{\text{ }2}^{3}=\partial
_{\theta }f\text{ }d\phi ,
\end{equation}
where the `prime' symbol denotes differentiation with respect to $r$. The
flat indices corresponding to $(t,$ $r,$ $\theta ,$ $\phi )$ are denoted by $%
(0,$ $1,$ $2,$ $3).$ Upon substituting for the gauge fields and the spin
connections in the Killing spinor equation (\ref{st}) we obtain

\begin{eqnarray}
\left[ \partial _{t}+\frac{1}{2}e^{V}\gamma _{0}\left( V^{\prime
}e^{-U}\gamma _{1}\pm \frac{iqk}{(d-3)N^{2}}\gamma _{23}+l\right) \right]
\epsilon &=&0,  \notag \\
\left[ \partial _{z_{\alpha _{i}}}+\frac{1}{2}e^{V}\gamma _{\alpha
_{i}}\left( V^{\prime }e^{-U}\gamma _{1}\pm \frac{iqk}{(d-3)N^{2}}\gamma
_{23}+l\right) \right] \epsilon &=&0,  \notag \\
\left[ \partial _{r}+\frac{e^{U}}{2}\gamma _{1}\left( \pm \frac{iqk}{%
(d-3)N^{2}}\gamma _{23}+{l}\right) \right] \epsilon &=&0,  \notag \\
\left[ \partial _{\theta }-{\frac{1}{2}}\left( e^{-U}N^{\prime }\gamma
_{1}-lN\right) \gamma _{2}\mp \frac{iqk}{2N}\gamma _{3}\right] \epsilon &=&0,
\notag \\
\left[ \partial _{\phi }-\frac{f}{2}\left( N^{\prime }e^{-U}\gamma _{1}+%
\frac{\partial _{\theta }f}{f}\gamma _{2}-lN\right) \gamma _{3}\pm \frac{ikqf%
}{2N}\gamma _{2}\mp \frac{i(d-2)}{2}lqk\int fd\theta \right] \epsilon &=&0.
\label{gen}
\end{eqnarray}

\subsection{Spherical and Hyperbolic Transverse Space}

We first consider magnetic branes where the transverse space is either
hyperbolic or spherical. The gauge field for these cases is given by $%
F_{\theta \phi }=\pm \kappa qf$ , ($\kappa =1,-1).$ Supersymmetric magnetic
string solutions of five dimensional $N=2$ supergravity theories were
considered in \cite{chamsabra, ks}. As in the five dimensional case, we
assume that the Killing spinors satisfy the following conditions:

\begin{equation}
\gamma _{2}\gamma _{3}\epsilon =\pm i\epsilon ,\ \ \ \gamma _{1}\epsilon
=-\epsilon .
\end{equation}
Then, from $\left( \ref{gen}\right) $ we obtain for $N(r)=r$: 
\begin{eqnarray}
\left[ \partial _{t}-\frac{1}{2}e^{V}\gamma _{0}\left( V^{\prime }e^{-U}+%
\frac{\kappa q}{(d-3)r^{2}}-l\right) \right] \epsilon  &=&0,  \notag \\
\left[ \partial _{z_{\alpha _{i}}}-\frac{1}{2}e^{V}\gamma _{\alpha
_{i}}\left( V^{\prime }e^{-U}+\frac{\kappa q}{(d-3)r^{2}}-l\right) \right]
\epsilon  &=&0,  \notag \\
\left[ \partial _{r}+\frac{e^{U}}{2}\left( \frac{\kappa q}{(d-3)r^{2}}-{l}%
\right) \right] \epsilon  &=&0,  \notag \\
\left[ \partial _{\theta }-{\frac{1}{2}}\left( e^{-U}-\frac{\kappa q}{r}%
-lr\right) \gamma _{2}\right] \epsilon  &=&0,  \notag \\
\left[ \partial _{\phi }\mp \frac{i}{2}\left( \partial _{\theta
}f+(d-2)lq\kappa \int fd\theta \right) -\frac{f}{2}\left( e^{-U}-\frac{%
\kappa q}{r}-lr\right) \gamma _{3}\right] \epsilon  &=&0.  \label{se}
\end{eqnarray}
The above equations imply the following conditions: 
\begin{eqnarray}
\partial _{t}\epsilon  &=&\partial _{z_{\alpha _{i}}}\epsilon =\partial
_{\theta }\epsilon =\partial _{\phi }\epsilon =0,  \notag \\
q &=&\frac{1}{(d-2)l},  \notag \\
e^{-U} &=&\frac{\kappa q}{r}+lr=\frac{\kappa }{(d-2)lr}+lr,  \notag \\
V^{\prime } &=&e^{U}\left( l-\frac{q\kappa }{(d-3)r^{2}}\right) .
\label{list}
\end{eqnarray}
The differential equation for $V,$ given by the last equation in (\ref{list}%
), can be easily solved and gives 
\begin{equation}
e^{V}=(lr)\left( 1+\frac{\kappa }{(d-2)l^{2}r^{2}}\right) ^{\frac{1}{2}%
\left( \frac{d-2}{d-3}\right) }.
\end{equation}
To summarize, we have obtained the solutions of \cite{waf} by seeking
magnetic brane solutions admitting Killing spinors. The dependence of the
magnetic charge on the cosmological constant found in \cite{waf} is
connected to the fact that the corresponding solutions admit Killing spinors.

From (\ref{list}), we find that the Killing spinors only depend on the
coordinate $r.$ \ Using the third equation in (\ref{se}) together with the
last equation in (\ref{list}), we obtain a simple differential equation for
the radial dependence of the Killing spinors, given by 
\begin{equation}
\left( \partial _{r}-{\frac{1}{2}}V^{\prime }\right) \epsilon =0,
\end{equation}
and hence 
\begin{equation}
\epsilon =e^{{\frac{1}{2}}V(r)}\epsilon _{0}=(lr)^{\frac{1}{2}}\left( 1+%
\frac{\kappa }{(d-2)l^{2}r^{2}}\right) ^{\frac{1}{4}\left( \frac{d-2}{d-3}%
\right) }\epsilon _{0},  \label{kill}
\end{equation}
where $\epsilon _{0}$ is a constant spinor satisfying the constraints $%
\gamma _{2}\gamma _{3}\epsilon _{0}=\pm i\epsilon _{0}$ and $\gamma
_{1}\epsilon _{0}=-\epsilon _{0}.$

It must be noted that, since the Killing spinors are independent of the
transverse coordinates $\theta $ and $\phi $, we could (for $\kappa =-1$)
compactify the $H^{2}$ to an arbitrary genus Riemann surface, and the
resulting solution will still have the Killing spinors (\ref{kill}). The
spherical branes contain a naked singularity but the hyperbolic ones have an
event horizon at $r=1/(l\sqrt{d-2})$. In the near horizon region, the
hyperbolic solution reduces to the product manifold $AdS_{(d-2)}\times H^{2}$%
.

\subsection{Flat Transverse Space}

Here we consider brane solutions with flat transverse space where the
magnetic field vanishes and the Killing spinor equations (\ref{gen}) yield

\begin{eqnarray}
\left[ \partial _{t}+\frac{1}{2}e^{V}\gamma _{0}\left( V^{\prime
}e^{-U}\gamma _{1}+l\right) \right] \epsilon &=&0,  \notag \\
\left[ \partial _{z_{\alpha _{i}}}+\frac{1}{2}e^{V}\gamma _{\alpha
_{i}}\left( V^{\prime }e^{-U}\gamma _{1}+l\right) \right] \epsilon &=&0, 
\notag \\
\left[ \partial _{r}+\frac{le^{U}}{2}\gamma _{1}\right] \epsilon &=&0, 
\notag \\
\left[ \partial _{\theta }+\frac{1}{2}\gamma _{2}(e^{-U}\gamma _{1}+lr)%
\right] \epsilon &=&0,  \notag \\
\left[ \partial _{\phi }+\frac{1}{2}\gamma _{3}(e^{-U}\gamma _{1}+lr)\right]
\epsilon &=&0.  \label{hai}
\end{eqnarray}
Integrability conditions of the above differential equations fix the metric
completely and give simply 
\begin{equation}
e^{V}=e^{-U}=lr.
\end{equation}
The solution for the Killing spinors is then given by

\begin{equation}
\epsilon =e^{-\frac{1}{2}V}\epsilon _{+}^{0}-e^{\frac{1}{2}V}\left( l\gamma
_{0}t+l\sum \gamma _{\alpha _{i}}z^{\alpha _{i}}+\gamma _{2}\theta +\gamma
_{3}\phi \right) \epsilon _{+}^{0}+e^{\frac{1}{2}V}\epsilon _{-}^{0}
\label{ni}
\end{equation}
where $\epsilon _{\pm }^{0}$ are constant spinors satisfying $\ (1\mp \gamma
_{1})\epsilon _{\pm }^{0}=0$. Clearly there are no constraints on the
Killing spinors and the solution is locally $AdS_{d}$. We may also consider
a quotient space of $AdS_{d}$ by compactifying the $(\theta ,\phi )$ sector
to a cylinder or a torus. Then, the surviving Killing spinors are those
which respect the identifications performed in the $(\theta ,\phi )$ sector.
These are given by 
\begin{equation}
\epsilon =e^{\frac{1}{2}V}\epsilon _{-}^{0}.
\end{equation}

\subsection{ Constant Warping Function}

Finally we set $N$ to a constant value $\mathcal{N}$ and seek magnetic brane
solutions admitting Killing spinors. In \cite{waf} it was shown that, for a
constant warp factor, the equations of motion are solved by a product of two
spaces, a $(d-2)$-dimensional space, $M_{d-2},$ and the two-manifold $%
\mathcal{S}$. For various topologies of $\mathcal{S},(k=0,\pm 1),$ the
magnetic charge satisfies

\begin{equation}
q^{2}=(d-1)l^{2}\mathcal{N}^{4}+k\mathcal{N}^{2}
\end{equation}
and the Ricci tensor of $M_{d-2}$ is given by \ \ \ \ 
\begin{equation}
R_{mn}=-g_{mn}\left[ \frac{(d-1)(d-2)}{\left( d-3\right) }l^{2}+\frac{k}{%
\left( d-3\right) \mathcal{N}^{2}}\right] .
\end{equation}
It can be shown that, for a constant warp factor, we can obtain magnetic
branes with Killing spinors only for $k=-1$. The gauge field is given by $%
F_{\theta \phi }=\mp qf$ and the Killing spinors satisfy the condition 
\begin{equation}
\gamma _{2}\gamma _{3}\epsilon =\pm i\epsilon .
\end{equation}
The integrability conditions for the Killing spinor equations in this case
imply the following relations: 
\begin{eqnarray}
q &=&\frac{1}{(d-2)l},\text{\ \ }  \notag \\
\mathcal{N}^{2} &=&\frac{1}{(d-2)l^{2}},  \notag \\
V^{\prime }e^{-U} &=&\frac{(d-2)}{(d-3)}l.  \label{tim}
\end{eqnarray}
The Killing spinor equations are finally given by

\begin{eqnarray}
\left[ \partial _{t}+le^{V}\frac{(d-2)}{2(d-3)}\gamma _{0}\left( 1+\gamma
_{1}\right) \right] \epsilon &=&0,  \notag \\
\left[ \partial _{z_{\alpha _{i}}}+le^{V}\frac{(d-2)}{2(d-3)}\gamma _{\alpha
_{i}}\left( 1+\gamma _{1}\right) \right] \epsilon &=&0,  \notag \\
\left[ \partial _{r}+le^{U}\frac{(d-2)}{2(d-3)}\gamma _{1}\right] \epsilon
&=&0,  \notag \\
\partial _{\theta }\epsilon &=&0,  \notag \\
\partial _{\phi }\epsilon &=&0.
\end{eqnarray}

Therefore, the configuration with a constant warp factor admitting Killing
spinors is $AdS_{d-2}\times H^{2}.$ These Killing spinors are given by 
\begin{equation}
\epsilon =e^{-\frac{1}{2}V}\epsilon _{+}^{0}-\frac{(d-2)}{(d-3)}le^{\frac{1}{%
2}V}\left( \gamma _{0}t+\sum \gamma _{\alpha _{i}}z^{\alpha _{i}}\right)
\epsilon _{+}^{0}+e^{\frac{1}{2}V}\epsilon _{-}^{0},  \label{kn}
\end{equation}
where $\epsilon _{\pm }^{0}$ are constant spinors satisfying 
\begin{eqnarray}
(1\mp \gamma _{1})\epsilon _{\pm }^{0} &=&0,\qquad  \notag \\
\gamma _{2}\gamma _{3}\epsilon _{\pm }^{0} &=&\pm i\epsilon _{\pm }^{0}.
\end{eqnarray}
Again, since the Killing spinors are independent of the transverse
coordinates $\theta $ and $\phi $, $H^{2}$ can be replaced by an arbitrary
genus Riemann surface and the resulting solutions will still have the
Killing spinors (\ref{kn}). The constant warp factor solution with Killing
spinors we found represents the near horizon geometry of the magnetic brane
with hyperbolic transverse space. We note that the Killing spinors of the
near horizon geometry of the hyperbolic solution satisfy only the condition $%
\gamma _{2}\gamma _{3}\epsilon =\pm i\epsilon $ whereas in the bulk they
satisfy an additional constraint, namely, $\gamma _{1}\epsilon =-\epsilon $.
This ``Killing spinor enhancement'' is analogous to the enhancement of
supersymmetry of the near horizon of BPS solutions of ungauged supergravity
theories \cite{enhance}.

\section{Curved Worldvolume Solutions and Killing Spinors}

In this section we generalize our magnetic brane solutions by allowing for a
general curved worldvolume. The gravitational equations of motion derived
from the action (\ref{action}) are given by

\begin{equation}
\hat{R}_{\mu \nu }=\frac{(d-2)}{(d-3)}F_{\mu q}F_{v}^{\text{ \ }q}-g_{\mu
\nu }\left( \frac{1}{2(d-3)}F_{pq}F^{pq}+(d-1)l^{2}\right) .  \label{me}
\end{equation}
\newline
We make the following ansatz for the curved $(d-4)$-magnetic brane: 
\begin{equation}
ds^{2}=e^{2V(r)}h_{mn}dx^{m}dx^{n}+e^{2U(r)}dr^{2}+r^{2}\left( d\theta
^{2}+f^{2}d\phi ^{2}\right) .
\end{equation}
where $h_{mn}$ is the metric of the $(d-3)$-dimensional worldvolume assumed
to depend only on the worldvolume coordinates $x^{m}$. For this ansatz we
obtain for the Ricci tensor: 
\begin{eqnarray}
\hat{R}_{mn} &=&R_{mn}-h_{mn}e^{2V-2U}\text{ }\left[ (d-3)V^{\prime
2}-U^{\prime }V^{\prime }+V^{\prime \prime }+\frac{2V^{\prime }}{r}\right] ,
\notag \\
\hat{R}_{rr} &=&-\left[ (d-3)\left( V^{\prime 2}-U^{\prime }V^{\prime
}+V^{\prime \prime }\right) -\frac{2U^{\prime }}{r}\right] ,  \notag \\
\hat{R}_{\theta \theta } &=&\frac{1}{f^{2}}\hat{R}_{\phi \phi }=-e^{-2U}%
\left[ \text{ }(d-3)rV^{\prime }-rU^{\prime }-ke^{2U}+1\right] ,
\end{eqnarray}
where $R_{mn}$ is the Ricci tensor of the worldvolume. The equations of
motion (\ref{me}) then give

\begin{eqnarray}
e^{-2V}R_{mn}-e^{-2U}h_{mn}\left[ (d-3)V^{\prime 2}-(U^{\prime }-\frac{2}{r}%
)V^{\prime }+V^{\prime \prime }\right] &=&-h_{mn}\left[ \frac{k^{2}q^{2}}{%
(d-3)r^{4}}+(d-1)l^{2}\right] ,  \notag \\
e^{-2U}\left[ (d-3)\left[ V^{\prime \prime }-V^{\prime }(U^{\prime
}-V^{\prime })\right] -\frac{2U^{\prime }}{r}\right] &=&\left[ \frac{%
k^{2}q^{2}}{(d-3)r^{4}}+(d-1)l^{2}\right] ,  \notag \\
e^{-2U}\left[ (3-d)rV^{\prime }+rU^{\prime }-1+ke^{2U}\right] &=&\left[ 
\frac{k^{2}q^{2}}{r^{2}}-r^{2}(d-1)l^{2}\right] ,
\end{eqnarray}
where we have used $F_{\theta \phi }=\pm kqf.$ Examining the above
equations, we find that the solutions 
\begin{equation*}
e^{V}=(lr)\left[ 1+\frac{k}{(d-2)l^{2}r^{2}}\right] ^{\frac{1}{2}\left( 
\frac{d-2}{d-3}\right) },\text{ \ \ \ \ \ }e^{U}=\left[ lr+\frac{k}{(d-2)lr}%
\text{\ }\right] ^{-1}\text{\ }
\end{equation*}
are still valid provided the worldvolume metric is Ricci-flat, i.e.,

\begin{equation}
R_{mn}=0.\ \ 
\end{equation}
For example, the four-dimensional Schwarzchild metric can be taken as the
worldvolume of a seven-dimensional magnetic brane with a flat, spherical or
hyperbolic transverse space. If we demand that our curved branes admit
Killing spinors, then the Killing spinor equation (\ref{st}) will imply, for 
$\gamma _{2}\gamma _{3}\epsilon =\pm i\epsilon \ $\ and $\ \gamma
_{1}\epsilon =-\epsilon ,$ that

\begin{equation}
\nabla _{m}\epsilon =0,  \label{paral}
\end{equation}
where $\nabla _{m}$ is the covariant derivative for the metric $h_{mn}$. The
dependence of the Killing spinors on the transverse coordinates is still the
same as in the Minkowski worldvolume case. Therefore, for the curved
magnetic branes to admit Killing spinors in $d$ dimensions with the
projection conditions as stated, their $\left( d-3\right) $-dimensional
worldvolume must admit parallel spinors.

As has been demonstrated in \cite{jose}, the integrability of (\ref{paral})
implies the following condition (no summation for $m$):

\begin{equation}
R_{mn}R_{mp}g^{np}\epsilon =0.  \label{sah}
\end{equation}

A manifold satisfying the condition (\ref{sah}) is called Ricci-null. A
Riemannian manifold has no null vectors, in which case (\ref{sah}) simply
implies that $R_{mn}=0$, i.e. Riemannian Ricci-null manifolds are Ricci-flat
and thus satisfy Einstein's equations of motion. However, in a Lorentzian
spacetime, the integrability condition (\ref{sah}) does not necessarily
imply Ricci-flatness. In searching for curved magnetic brane solutions with
Killing spinors, we may consider two types of worldvolume metrics with
parallel spinors, static metrics with a covariantly constant time-like
vector and nonstatic metrics with covariantly constant light-like vector.
The metric of a static worldvolume with parallel spinors is given by

\begin{equation}
ds^{2}=-dt^{2}+ds^{2}(X)
\end{equation}
where $X$ is a Ricci-flat Riemannian manifold admitting parallel spinors$.$
The number of parallel spinors depends on the holonomy group of the manifold 
$X$ (see \cite{perry, jose} for a discussion). For a nonstatic worldvolume,
we consider a gravitational $pp$ wave which by definition is a spacetime
admitting a parallel null vector. General $d$-dimensional Lorentzian metric
admitting a parallel null vector was given many years ago by Brinkmann \cite
{Brinkmann2}. This metric can be written in the form 
\begin{equation}
ds^{2}=2dx^{+}dx^{-}+a(dx^{+})^{2}+b_{i}dx^{i}dx^{+}+g_{ij}dx^{i}dx^{j}
\label{brin}
\end{equation}
where $i=1,...,d-2$, and $\partial _{-}a=0,$ $\partial _{-}b_{i}=0,$ $%
\partial _{-}g_{ij}=0.$ Possible holonomy groups of supersymmetric Brinkmann
waves for $d\leq 11$ as well as explicit Ricci-flat supersymmetric Brinkmann
waves and warped product solutions were given in \cite{jose} to which we
refer the reader for extensive details. As an example, we consider the seven
dimensional magnetic brane where the worldvolume is given by a
four-dimensional spacetime. Four-dimensional static Ricci-flat metrics are
automatically flat and therefore one needs to consider nonstatic metrics of
the form (\ref{brin}) to get curved worldvolumes. In this case, the holonomy
of the metric must be $\mathbb{R}^{2}\subset $ \ Spin(3,1) for it to admit a
parallel spinor. The metric with this holonomy group was constructed in \cite
{jose} and is given by

\begin{equation}
ds^{2}=2dx^{+}dx^{-}+a(dx^{+})^{2}+b\epsilon
_{ij}x^{j}dx^{i}dx^{+}+dx^{i}dx^{i},\text{ \ \ \ \ \ for \ }i,j=1,2
\end{equation}
where $b=b(x^{+})$ and $a=a(x^{+},x^{i})$ are arbitrary functions. For the
equations of motion to be satisfied one must choose $b^{2}=\frac{1}{4}\Delta
a,$ where $\Delta $ is the Laplacian on functions of $x^{i}.$

\section{\protect\bigskip Conclusions}

We have studied the Killing spinors of magnetic brane solutions of the $d$%
-dimensional AdS Einstein-Maxwell theories. The magnetic charge of these
solutions is related to the inverse of the cosmological constant. We also
generalized the solutions of \cite{waf} to include Ricci-flat worldvolumes.
Thus Ricci-flat solution of $(d-3)$-dimensional Einstein gravity can be
embedded in a magnetic brane solution of $d$-dimensional AdS
Einstein-Maxwell theory. If we demand that the $d$-dimensional curved
solution admits Killing spinors, then the ($d-3)$-dimensional Ricci-flat
worldvolume must admit parallel spinors. To construct worldvolumes with
parallel spinors one can consider a class of solutions which are locally
isometric to a product $\mathbb{R}$ $\times $ $X,$ with $X$ being a
Ricci-flat Riemannian manifold admitting parallel spinor. The other class of
worldvolumes with Killing spinors that one can consider are those given by
indecomposable Lorentzian spacetimes with a covariantly constant light-like
vector. These spacetimes are a subclass of Brinkmann waves. Restricting the
holonomy of these spaces in order to have parallel spinors in not enough to
ensure Ricci-flatness and therefore the equations of motion impose extra
conditions.

\bigskip

\textbf{Acknowledgements}

I would like to thank N. N. Khuri and his group at the Rockefeller
University for their hospitality during the preparation of this work.

\end{document}